\documentclass{PoS}
\usepackage{amsmath,amssymb,url}

\newcommand{\td}[3]{\frac{d^{#3} #1}{d {#2}^{#3}}} 
\renewcommand{\v}[1]{\ensuremath{\mathbf{#1}}} 

\title{Dark matter gets DAMPE at high energies}

\ShortTitle{Dark matter gets DAMPE}

\author{\speaker{Geoff Beck}\\
	School of Physics, University of the Witwatersrand, Private Bag 3, WITS-2050, Johannesburg, South Africa\\
	E-mail: \email{geoffrey.beck@wits.ac.za}}

\author{Sergio Colafrancesco\footnote{In honour of the memory of my colleague and mentor, who passed before this work was complete}\\
	School of Physics, University of the Witwatersrand, Private Bag 3, WITS-2050, Johannesburg, South Africa\\
	E-mail: \email{sergio.colafrancesco@wits.ac.za}}

\abstract{The DArk Matter Particle Explorer (DAMPE) mission revealed a break in the spectrum of cosmic-ray electons and positrons. This is associated with an excess above the expected backgrounds at energies around $1$ TeV. Several authors have argued that such an excess can be explained in terms of dark matter models that feature heavy leptophilic WIMPs. These models, however, require some form annihilation enchancement above that expected from the Milky-Way galactic centre. This can take the form of either a local over-density near to our solar system or some form of Sommerfeld enhancement of the annihilation rate. In this work we will explore the detectability of local over-densities using gamma-ray and neutrino observatories. We conclude that KM3NET may be the only up-coming high-energy instrument capable of ruling out the presence of such objects. However, in the case where the local over-density is an Ultra-Compact Mini Halo, CTA can also explore the parameter space of these proposed dark matter models. }

\FullConference{High Energy Astrophysics in Southern Africa - HEASA2018\\
		1-3 August, 2018\\
		Parys, Free State, South Africa}

\begin{document}
	
\section{Introduction}
\label{sec:intro}

In late 2017 the DArk Matter Particle Explorer (DAMPE) announced the detection of a break in the spectrum of cosmic-ray electrons/positrons~\cite{dampe}. This spectral break was accompanied by a significant excess of electrons/positrons above the expected backgrounds at energies around 1 TeV. This excess has been the source of some speculation in the literature with several Dark Matter (DM) models proposed to account for it \cite{dampedm1,dampedm2,dampeucmh,dampedm3,dampedm4}. What these models have in common is that they feature a large mass $m_{\psi} \gtrsim 1$ TeV WIMP particle $\psi$ which interacts with leptons in the Standard Model via a mediator boson of larger mass than the WIMP. Another commonality is necessity of some form of annihilation enhancement in order to simultaneously explain the excess and satisfy relic population limits. The works cited above utilise a local over-dense sub-halo of DM to produce this boost in annihilation rates. It has already been shown that such a local sub-halo would not be visible to Fermi-LAT \cite{dampedm1,dampedm2}. However, it may be possible to rule out the presence of such an object by other means. In particular we will determine in this work whether the upcoming Cherenkhov Telescope Array (CTA) or the KM3NET neutrino telescope will be capable of ruling out the presence of a local DM sub-halo across the allowed parameter space of models designed to explain the DAMPE excess. This work forms a complement to earlier work by the same authors where we studied the DAMPE parameter space using radio observations of target DM halos~\cite{gs-saip2018}.

We demonstrate that KM3NET shows the potential to probe a  large region of the allowed parameter space, possibly ruling out the presence of the sub-halo needed to explain the excess. This is based on preliminary estimates for the sensitivity of KM3NET to extended sources including only muon neutrino detection, these sensitivities are expected to improve with the inclusion of the other two species of neutrino~\cite{Ambrogi:2018skq}. Based on the same work we use CTA sensitivity to extended sources to show that the CTA will be largely unable to probe the DAMPE excess parameter space. With the exception of when the local sub-halo is an exotic object like an Ultra-Compact Mini Halo (UCMH) as suggested in \cite{dampeucmh}.   

We also study whether any existing or potential non-observation constraints can probe the DAMPE parameter space. We show that KM3NET may have a limited ability to do so with the galactic centre as a target source. 

This work is structured as follows: in section~\ref{sec:dampe} we elaborate on the DAMPE excess DM models we will study. In section~\ref{sec:dmann} we detail the annihilation formalism employed here and the emissions produced in section~\ref{sec:emm}. The results are presented in section~\ref{sec:res} and are discussed in \ref{sec:disc}.

\section{Dark Matter Models for the DAMPE Excess}
\label{sec:dampe}

The DM models considered are heavy leptophilic WIMPs $\psi$ that couple to the Standard Model particles via a heavy mediator that is too large to allow for the decay of the WIMP~\cite{dampedm1,dampedm2,dampedm4}. Hence only annihilation will be considered here. We will consider the following ranges from the models listed above: $\psi$ couples to muons and electrons and spans a mass range around $1.4$ to $1.7$ TeV with cross-sections ranging from $3\times 10^{-26}$ to $5\times 10^{-24}$ cm$^3$ s$^{-1}$ in accordance with \cite{dampedm1}. The emissions stem from a DM clump of mass $10^6$ M$_{\odot}$ within a distance of $0.1$ kpc \cite{dampedm1} or a Ultra-Compact Mini-Halo (UCMH) of mass $\sim 3$ M$_{\odot}$ within a distance of $0.3$ kpc~\cite{dampeucmh}. For details of the UCMH formalism we refer the reader to \cite{ricotti2009,bringmann2012}. 

The second set of models considered has $\langle \sigma V \rangle  = 3 \times 10^{-26}$ cm$^3$ s$^{-1}$ with the electron only coupling ($e^+e^-$) and three lepton democratic coupling ($3l$) cases. For the $3l$ case we will work in the scenario of a DM clump situated at $0.3$ kpc with a mass of $2\times 10^8$ M$_{\odot}$. For the case of coupling to electrons only we use a halo with mass $8.0\times 10^{7}$ M$_{\odot}$ within a distance $0.3$ kpc. 

The distance and mass choices are representative of the models as a whole, as the distance and mass must co-vary to maintain the same flux in accounting for the excess observed by DAMPE. Non-UCMH clumps are considered to have Navarro-Frenk-White (NFW)~\cite{nfw1996} density profiles with concentration parameters calculated according to \cite{prada2012}.

\section{Dark Matter Annihilation}
\label{sec:dmann}

The source function annihilation of WIMPs $\psi$ into final-state photons/neutrinos with energy $E$ at halo position $r$ is given by
\begin{equation}
Q_i (r,E) = \langle \sigma V\rangle \sum\limits_{f}^{} \td{N^f_i}{E}{} B_f \left(\frac{\rho_{\psi}(r)}{m_{\psi}}\right)^2 \; ,
\end{equation}
where $i \in \{\gamma,\,\nu\}$, $\langle \sigma V\rangle$ is the non-relativistic velocity-averaged annihilation cross-section at $0$ K, $B_f$ is the branching fraction for intermediate state $f$, $\td{N^f_i}{E}{}$ is the differential photon/neutrino yield of the $f$ channel, and $\left(\frac{\rho_{\psi}(r)}{m_{\psi}}\right)^2$ is the number density of pairs of WIMPs.

The functions $\td{N^f_i}{E}{}$ will be sourced from \cite{ppdmcb1,ppdmcb2}. We will follow the standard practice of studying each annihilation channel $f$ independently, assuming $B_f = 1$ for each separate case (an exception is the $3l$ case where we weight each lepton channel equally). The studied channels will all be leptonic: $\tau$ leptons, muons, and electrons/positrons in accordance with \cite{dampedm1,dampedm2,dampeucmh}.


\section{Gamma-ray and Neutrino Emission}
\label{sec:emm}
For the DM-induced $\gamma$-ray or neutrino production, the resulting flux calculation takes the form
\begin{equation}
S_{i} (E,z) = \int_0^r d^3r^{\prime} \, \frac{Q_{i}(E,z,r^{\prime})}{4\pi D_L^2} \; ,
\end{equation}
with $Q_{i}(\nu,z,r)$ being the source function for energy $E$ and position $r$ within the given DM halo at redshift $z$, and $D_L$ is the luminosity distance to the halo.
The spatial integration over the source function $Q$ will be summarised in the astrophysical J-factor of the target halo:
\begin{equation}
J (\Delta \Omega, l) = \int_{\Delta \Omega}\int_{l} \rho^2 (\v{r}^{\prime}) dl^{\prime}d\Omega^{\prime} \; , \label{eq:jfactor}
\end{equation}
with $\rho (r)$ being the halo density profile, the integral being extended over the line of sight $l$, and $\Delta \Omega$ is the observed solid angle.
The flux can then be written as
\begin{equation}
S_{i} (E,z) = \langle \sigma V\rangle \sum\limits_{f}^{} \td{N^f_i}{E}{} B_f J(\Delta \Omega,l) \; .
\end{equation}

\section{Results}
\label{sec:res}

Here we present the results of calculating both the impact of existing limits from Fermi-LAT~\cite{Fermidwarves2015} and projected limits for CTA and KM3NET. There are two approaches used. In the first we determine which cross-section values may be ruled out through non-observation of a local sub-halo using either CTA or KM3NET. In the second approach we determine how much of the parameter space could be constrained through non-observation of neutrino fluxes from the galactic centre (with halo parameters from \cite{fermigc2015} ). For all these applications we use extended source sensitivities as calculated by \cite{Ambrogi:2018skq} (see also \url{http://www.cta-observatory.org/science/cta-performance/}). 

\begin{figure}[htbp]
	\centering
	\includegraphics[scale=0.6]{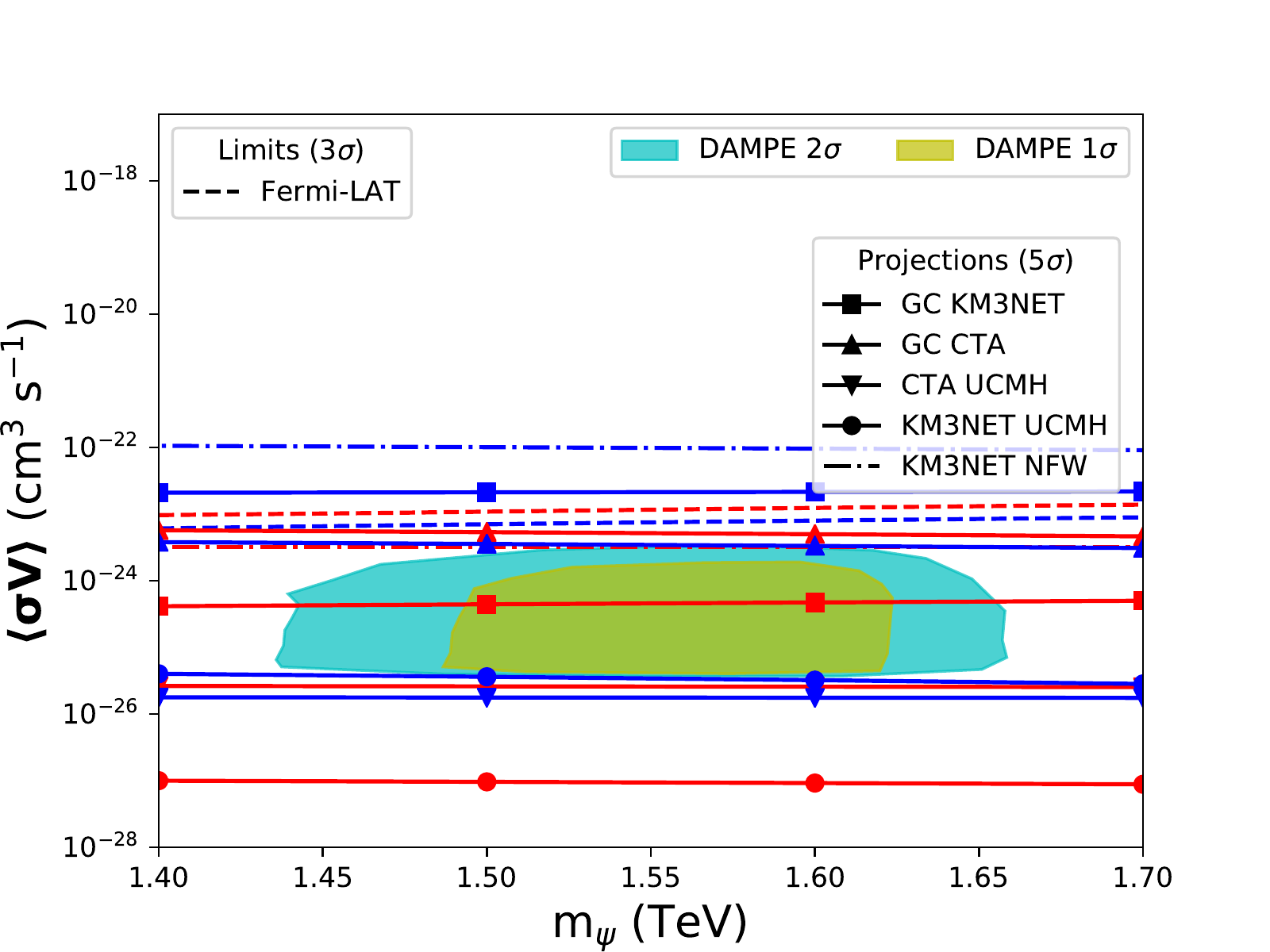}
	\caption{Parameter space for models from \cite{dampedm1} and \cite{dampeucmh}. Lines labelled with GC refer to galactic centre projections while UCMH and NFW lines refer to direct sub-halo searches with UCMH or NFW profiles respectively. Red lines display coupling to muons only while blue lines show those for electrons.}
	\label{fig:dampe1}
\end{figure}

In figure~\ref{fig:dampe1} we display results super-imposed on the contours from \cite{dampedm1}, these take into account direct detection, CMB constraints, the DAMPE excess, and thermal relic population limits. Both CTA and KM3NET can rule out the presence of the sub-halo when it is in the form of a UCMH for both electron and muon couplings. When the object is a more extended NFW~\cite{nfw1996} sub-halo we find that neither CTA nor KM3NET can provide any constraints. Projected limits from non-observation of galactic centre (labelled with GC in the figure) neutrino flux with KM3NET are able to probe about half-way into the parameter space, but only in the case of muon coupling. Best-case CTA observations of the galactic centre cannot provide meaningful constraints.

In the case of the democratic $3$-lepton model or the electron-only case from \cite{dampedm2} (which have only a single provided cross-section value) we cannot probe down to the relic level in any case (barring the unrealistic point-source (PS) CTA projections). 

\begin{figure}[htbp]
	\centering
	\includegraphics[scale=0.6]{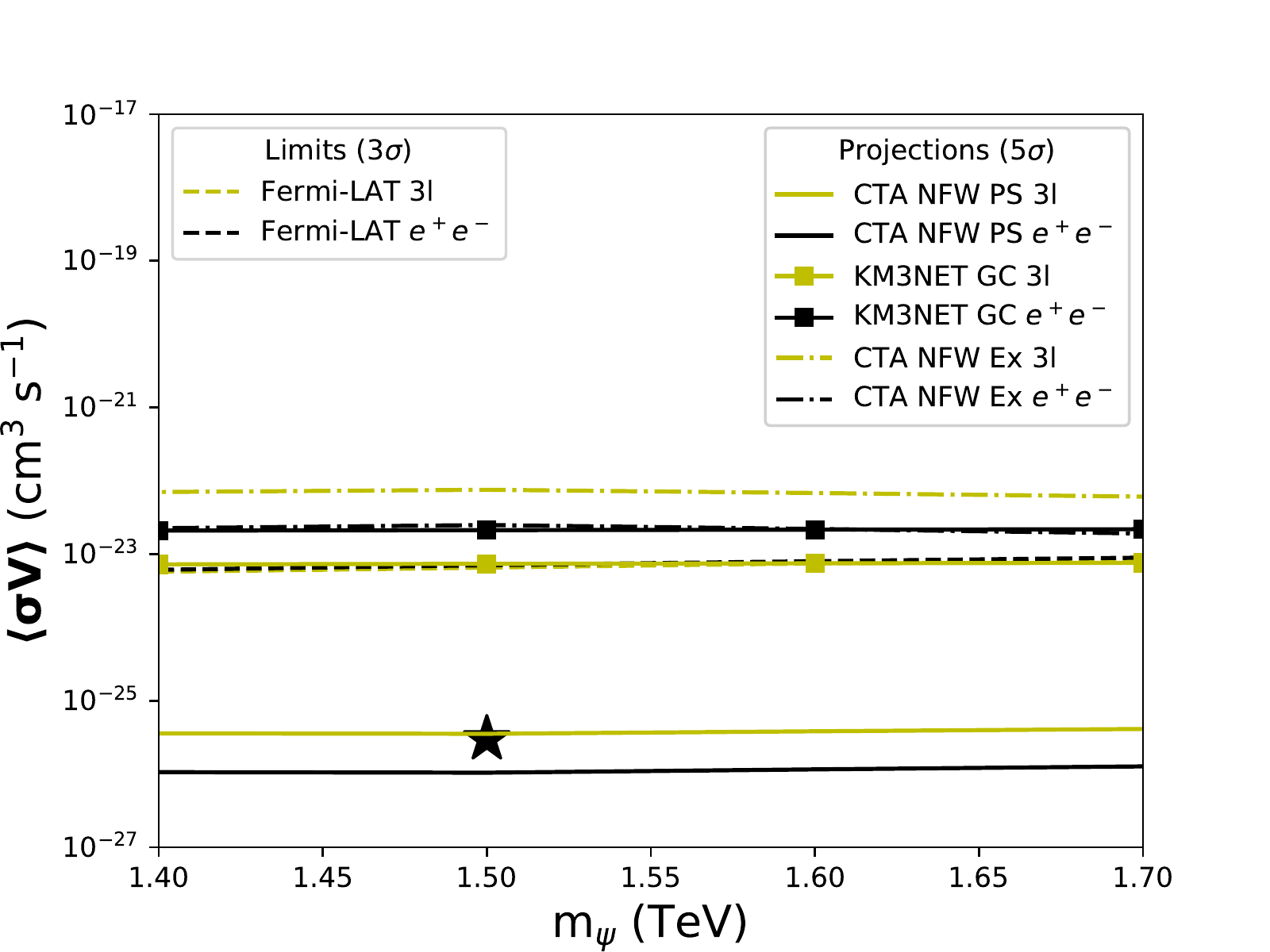}
	\caption{Parameter space for models from \cite{dampedm2}. Lines labelled with GC refer to galactic centre projections while the displayed CTA lines are for point-source (PS) and extended (Ex) sensitivity searches for the sub-halo. Black lines show the electron only case while yellow lines show the $3$-lepton case (both models from \cite{dampedm2}).}
\end{figure}

\section{Discussion and Conclusions}
\label{sec:disc}

In previous work~\cite{gs-saip2018} these authors had shown that muon neutrino fluxes inferred from galactic centre gamma-ray fluxes following \cite{gcneutrino} could place limits upon muon coupling of the proposed DAMPE models. In addition to this, the SKA was shown to be able to probe the entire parameter space in hunting the local sub-halo, even when making some accounting for the angular extension of the object. Here we show that even the upcoming CTA is unable to rule out a local sub-halo unless it has an ultra-compact density profile, this is largely due to the comparative angular extension of less exotic density profiles. However, it is established that CTA is substantially better at probing large mass WIMP models than Fermi-LAT. In the case of KM3NET, even with the conservative sensitivity employed (it considers only muon neutrinos), we find that it can also only detect an ultra-compact local sub-halo. Despite this, KM3NET could probe about half of the suggested DAMPE parameter space (for muon couplings only) via non-observation constraints on an extended-source flux from the Milky-Way galactic centre.

Thus, we have demonstrated the difficulty in probing DM models suggested to explain the DAMPE excess. This is despite the presence of a local over-dense sub-halo which enhances the annihilation rate of DM near the solar system. In comparison with \cite{gs-saip2018} we have established that the most promising strategy for probing DAMPE excess DM models is the use of up-coming radio and neutrino experiments. The gamma-ray options that have been explored are less promising for direct over-density detection, but, a multi-messenger strategy combining high and low energy observations is still available. Additionally, the potential of other targets, like dwarf galaxies, in high-energy DAMPE constraints will be explored in future work.

\section*{Acknowledgements}
This research has made use of the CTA instrument response functions provided by the CTA Consortium and Observatory, see \url{http://www.cta-observatory.org/science/cta-performance/} (version prod3b-v1) for more details.

\bibliographystyle{unsrt}
\bibliography{dampe_II.bib}

\begin{thebibliography}{10}

\bibitem{dampe}
G.~Ambrosi et~al.
\newblock {\em Nature}, 552:63, 2017.

\bibitem{dampedm1}
Y.~Fan, W.~Huang, M.~Spinrath, Y.~S. Tsai, and Q.~Yuan.
\newblock {\em Physics Letters}, B 781, 2018.

\bibitem{dampedm2}
Q.~Yuan et~al.
\newblock {\em Preprint}, page arXiv: 1711.10989, 2017.

\bibitem{dampeucmh}
F.~Yang, M.~Su, and Y.~Zhao.
\newblock {\em Preprint}, page arXiv: 1712.01724, 2017.

\bibitem{dampedm3}
Junjie Cao, Lei Feng, Xiaofei Guo, Liangliang Shang, Fei Wang, and Peiwen Wu.
\newblock Scalar dark matter interpretation of the dampe data with u(1) gauge
  interactions.
\newblock {\em Phys. Rev. D}, 97:095011, May 2018.

\bibitem{dampedm4}
Peter Athron, Csaba Balazs, Andrew Fowlie, and Yang Zhang.
\newblock Model-independent analysis of the dampe excess.
\newblock {\em Journal of High Energy Physics}, 2018(2):121, Feb 2018.

\bibitem{gs-saip2018}
G.~Beck and S.~Colafrancesco.
\newblock {\em Preprint}, page arXiv:1810.07176, 2018.

\bibitem{Ambrogi:2018skq}
Lucia Ambrogi, Silvia Celli, and Felix Aharonian.
\newblock {On the potential of Cherenkov Telescope Arrays and KM3 Neutrino
  Telescopes for the detection of extended sources}.
\newblock {\em Astropart. Phys.}, 100:69--79, 2018.

\bibitem{ricotti2009}
M.~Ricotti and A.~Gould.
\newblock {\em Astrophys. J.}, 707:979, 2009.

\bibitem{bringmann2012}
T.~Bringmann, P.~Scott, and Y.~Akrami.
\newblock {\em Phys. Rev.}, D85:125027, 2012.

\bibitem{nfw1996}
J.~F. Navarro, C.~S. Frenk, and S.~D.~M. White.
\newblock {\em Astrophys. J.}, 462:563, 1996.

\bibitem{prada2012}
F.~Prada, A.~A. Klypin, A.~J. Cuesta, J.~E. Betancort-Rijo, and J.~Primack.
\newblock {\em MNRAS}, 423(4):3018, 2012.

\bibitem{ppdmcb1}
M.~Cirelli et~al.
\newblock {\em JCAP}, 1103:051, 2011.

\bibitem{ppdmcb2}
P.~Ciafaloni et~al.
\newblock {\em JCAP}, 1103:019, 2011.

\bibitem{Fermidwarves2015}
A.~Drlica-Wagner et~al.
\newblock {\em Astrophys. J.}, 809:L4, 2015.

\bibitem{fermigc2015}
M.~Ajello et~al.
\newblock {\em Astrophys. J.}, 819(1):44, 2016.

\bibitem{gcneutrino}
S.~Celli, A.~Palladino, and F.~Vissani.
\newblock {\em Eur. Phys. J.}, C77:66, 2017.

\end{thebibliography}

\end{document}